\documentclass[twocolumn,english,aps,prb,floatfix,preprintnumbers,showpacs,amsfonts,amssymb,superscriptaddress]{revtex4}
\usepackage[T1]{fontenc}
\usepackage[latin1]{inputenc}
\usepackage{amsmath}
\usepackage{graphicx}
\usepackage{amssymb}

\makeatletter
\usepackage{amscd}
\usepackage{bm}

\usepackage{babel}
\makeatother
\begin{document}

\title{Phase Diagram of the Two-Leg Kondo Ladder}

\author{J.~C.~Xavier}

\affiliation{Instituto de F\'{\i}sica Gleb Wataghin, Unicamp, Caixa Postal 6165, Campinas
SP 13083-970, Brazil}

\author{E.~Miranda}

\affiliation{Instituto de F\'{\i}sica Gleb Wataghin, Unicamp, Caixa Postal 6165, Campinas
SP 13083-970, Brazil}

\author{E.~Dagotto}

\affiliation{National High Magnetic Field Lab and Department of Physics,~Florida~State~University,~Tallahassee,~Florida~32306 }

\date{\today{}}

\begin{abstract}
The phase diagram of the two-leg Kondo ladder is investigated using computational 
techniques. Ferromagnetism is present, but only at small conduction electron densities 
and robust Kondo
coupling $J$. For densities $n\gtrsim0.4$ and any Kondo coupling,
a paramagnetic phase is found. We also observed spin dimerization
at densities $n$=$1/4$ and $n$=$1/2$. The spin structure
factor at small $J$ peaks at $\vec{q}$=$(2n,0)\pi$
for $n\lesssim0.5$, and at $\vec{q}$=$(n,1)\pi$ for $n\gtrsim0.5$.
The charge structure factor suggests that electrons behave as
free particles with spin-1/2 (spin-0) for small (large) $J$.
\end{abstract}
\maketitle
Numerical studies provide valuable unbiased
information about strongly correlated electronic systems.
However, current computer limitations restrict most investigations
to one-dimensional (1D) models or small two-dimensional (2D) clusters. 
Unfortunately, the physics found in 1D models is often qualitatively different
from results observed in real materials. A possible
procedure to start an investigation of realistic 2D models is by
coupling 1D systems together. This approach has been used with
great success in the case of the $t$-$J$ model, where numerical methods
such as Exact Diagonalization\cite{dagottorev} and Density Matrix
Renormalization Group (DMRG)\cite{white} were helpful
in elucidating several properties of the 2D model through studies
of $t$-$J$ ladders.\cite{dagottosc1,dagottorev,noacketal,whiteetal}

The ground-state properties of the 2D Kondo lattice model
(KLM) remain mainly unexplored using unbiased methods. 
The KLM is the simplest model believed to describe
heavy-fermion materials\cite{hewson} and, thus, a better understanding
of its ground state is much needed. The main goal of the
present work is to provide the first steps toward determining
the phase diagram of the two-leg Kondo ladder (2-LKL). 
The information provided here
will hopefully be as relevant for KLM 2D systems as ladder studies were in
the $t$-$J$ model context. The $N$-leg
Kondo ladders consist of $N$ Kondo chains coupled by the hopping
term. As shown below, the phase diagram of the 2-LKL is fairly different
from the 1D KLM.
We considered the 2-LKL on $2$$\times$$L$ clusters and a Hamiltonian
\[
H=-t \sum_{<i,j>,\sigma}(c_{i,\sigma}^{\dagger}c_{j,\sigma}^{\phantom{\dagger}}+\mathrm{H.}\,\mathrm{c.})+J\sum_{j}\mathbf{S}_{j}\cdot\mathbf{s}_{j},\]
 where $c_{j\sigma}$ annihilates a conduction electron at site $j$
with spin projection $\sigma$, $\mathbf{S}_{j}$ is a localized 
spin-$\frac{1}{2}$ operator, 
$\mathbf{s}_{j}$=$\frac{1}{2}\sum_{\alpha\beta}c_{j,\alpha}^{\dagger}\bm{\sigma}_{\alpha\beta}c_{j,\beta}^{\phantom{\dagger}}$
is the conduction electron spin density operator, and $\bm{\sigma}_{\alpha\beta}$
are Pauli matrices. Here $\langle ij \rangle$ denotes nearest-neighbor sites, $J$$>$$0$
is the Kondo coupling constant between conduction electrons and
local moments, and $t$=1 fixes
the energy scale. The total number of conduction electrons is $N$
and $n$=$N/2L$. This model was investigated with the DMRG technique\cite{white}
using open boundary conditions. The finite-size algorithm for
sizes up to $2$$\times$$L$=$80$ was applied, keeping up to $m$=$1200$ states per block.
The discarded weight was typically about $10^{-5}$-$10^{-9}$ in the
final sweep.

Let us first briefly describe what is currently known about the KLM
ground-state phase diagram. In 1D, for low electronic
density and/or large $J$ the ground state is ferromagnetic \cite{Tsunetsugu}
(see also Ref.~\onlinecite{ferrogu}).
The rest of the phase diagram is characterized by a paramagnetic phase,
except for a small wedge of ferromagnetism for fillings 
above $n$=$0.5$.\cite{mccullochetal}
Furthermore, spin dimerization has recently been discovered 
at $n$=$0.5$.\cite{dimer} Most ground-state investigations 
in higher dimensions have been limited to approximate approaches.
Doniach\cite{Doniach} pointed out the possible existence of
a Kondo-lattice
quantum critical point (QCP) due to the competition
between the Ruderman-Kittel-Kasuya-Yosida (RKKY) interaction, which
favors antiferromagnetism (AFM), 
and the Kondo effect, which favors paramagnetism.
The full mean-field phase diagram of the 3D Kondo lattice was
obtained by Lacroix and Cyrot.\cite{lacroixcyrot} They found that,
at small $J$, there is a critical density $n_{\rm c}$ separating
a ferromagnetic phase from an antiferromagnetic one. For sufficiently
large $J$, however, the Kondo effect dominates and the system is
paramagnetic. Further studies also considered an explicit exchange interaction
between localized spins.\cite{iglesias} Recently, quantum
Monte Carlo (QMC)\cite{assaad1} and DMRG\cite{xavier2} investigations
of the half-filled Kondo lattice in small clusters confirmed the existence
of a QCP at $J$$\sim$$1.45$ in agreement with previous 
approximate approaches.\cite{shietal}
Moreover, DMRG results on $N$-leg Kondo ladders at half-filling have
shown that the spin and charge gaps are nonzero for any number of
legs and coupling $J$.\cite{xavier2} A two-channel version
has also been studied at half-filling.\cite{morenoetal}
Here, the two-leg Kondo ladder $away$ from half-filling is considered.

\begin{figure}[htbp]
\begin{center}\includegraphics[%
  scale=0.27]{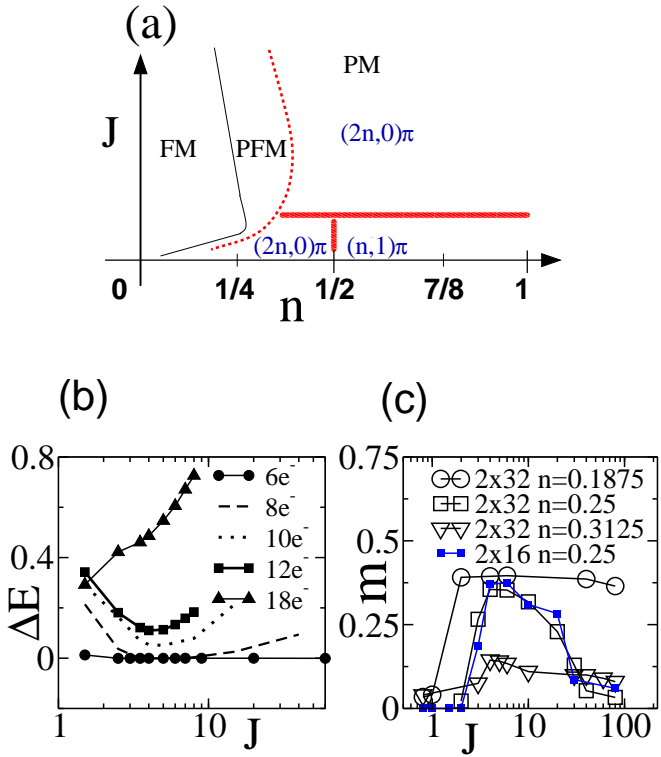}\end{center}

\begin{center}\includegraphics[%
  scale=0.31]{fig1bc.eps}\end{center}

\caption{\label{fig1} (a) Phase diagram of the 2-LKL. FM, PM, and PFM denote
regions with ferromagnetism, paramagnetism, and partial ferromagnetism,
respectively. To the right of the dotted line, the thick lines separate
three regions: small and large $J$ and $n\gtrless0.5$. These regions
are characterized by the location of the spin-spin structure
factor peak (see text). (b) Gap $\Delta E$ between the ground state
and the ferromagnetic state vs. $J$ for a 2$\times$16 cluster, and several
numbers of electrons (as shown). The densities from the top are,
$n$=$0.1875,$ 0.25, 0.3125, 0.374, and 0.5625. (c) Magnetization density
vs. Kondo coupling, at several densities $n$. The errors
are of the order of or smaller than the symbol size.}
\end{figure}

\begin{figure}[htbp]
\begin{center}\includegraphics[%
  scale=0.31]{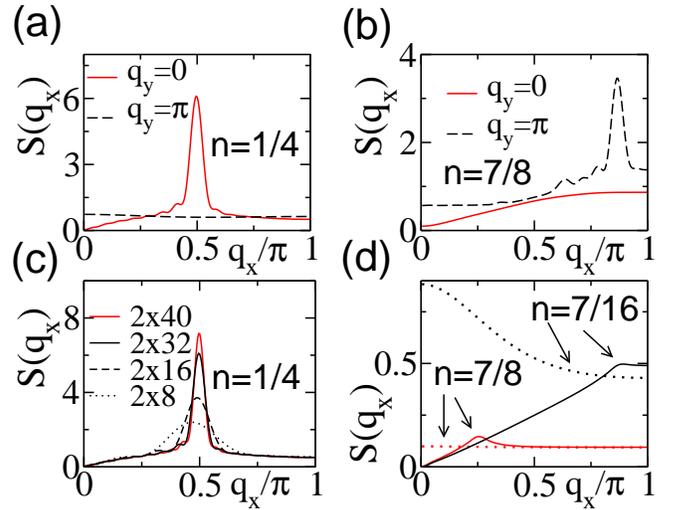}\end{center}

\caption{\label{fig2} The spin structure factor $S\left(\vec{q}\right)$
vs. $q_{x}$ for the 2-LKL: (a) $J$=$0.8$, $n$=$1/4$ and $L$=$32$; (b)
$J$=$0.8$, $n$=$7/8$ and $L$=$32$; (c) $S\left(q_{x},q_{y}=0\right)$
vs. $q_{x}$ for several values of $L$ with $J$=$0.8$ and $n$=$1/4$;
(d) $S\left(\vec{q}\right)$ for densities $n$=$7/16$ and $n$=$7/8$
(see arrows) with $L$=$32$ and $J$=$60$. The solid (dotted) lines correspond
to $q_{y}$=$0$ ($q_{y}$=$\pi$).}
\end{figure}

In Fig.~\ref{fig1}(a), the schematic phase diagram of the 2-LKL is
presented.  We have identified three phases characterized by full
ferromagnetism (FM), partially saturated FM (PFM), and paramagnetism
(PM). The approximate boundaries between these phases were first
obtained from the energy difference $\Delta E$=$E((2L-N)/2)-E(0)$ for
$L=16$, where $E(p)$ is the ground state energy in the sector with
total spin projection $S_{\rm
T}^{z}$=$\sum_{i}S_{i}^{z}+s_{i}^{z}$=$p$. This was then
double-checked through the (computationally more costly) calculation
of the ground-state total spin of 2$\times$8, 2$\times$16, and
2$\times$32 clusters. In Fig.~\ref{fig1}(b), we show $\Delta E$ as a
function of $J$ for the 2$\times$16 cluster, and several values of
$N$. These values suggest that for small density and large (small) $J$
the ground state is FM (PM),\cite{comme} while for the whole region
with density $n\gtrsim0.4$ and $J$$>$$0$ it is PM. In
Fig.~\ref{fig1}(c), the magnetization density $m$=$S_{\rm T}/2L$
vs. $J$ is shown for some conduction electron densities.  It can be
seen that for small $n$ and $J$ the total spin is zero (within the
DMRG precision). At $n$=$1/8$, $m$ starts to increase as $J$ is
increased, and saturates at $m$=$\left(1-n\right)/2$. For $n$=$1/4$,
however, $m$ is non-monotonic and vanishes with further increase of
$J$, apparently continuously. These results are not due to finite-size
effects: the same behavior is observed for both 2$\times$16 and
2$\times$32 clusters at this density, suggesting that it survives the
thermodynamic limit. For $0.25\lesssim n\lesssim0.4$, the
magnetization density does not saturate at $m$=$\left(1-n\right)/2$
(see $n$=$0.3125$, Fig.~\ref{fig1} (c)). Then, this phase has partial
ferromagnetic (PFM) order. For $n\gtrsim0.4$, we have found that
$m$$<$$0.03$ for several Kondo couplings $J$ (while for large $J$,
$m$$<$$10^{-3}$).  This result strongly suggests that the whole region
with $n\gtrsim0.4$ is paramagnetic.  This is different from the 1D
KLM, which shows full FM at any density for large enough
$J$.\cite{Tsunetsugu,mccullochetal} It is surprising that coupling
just one more chain to the 1D Kondo lattice induces such a dramatic
effect on the phase diagram.

\begin{figure}[htbp]
\begin{center}\includegraphics[%
  scale=0.13]{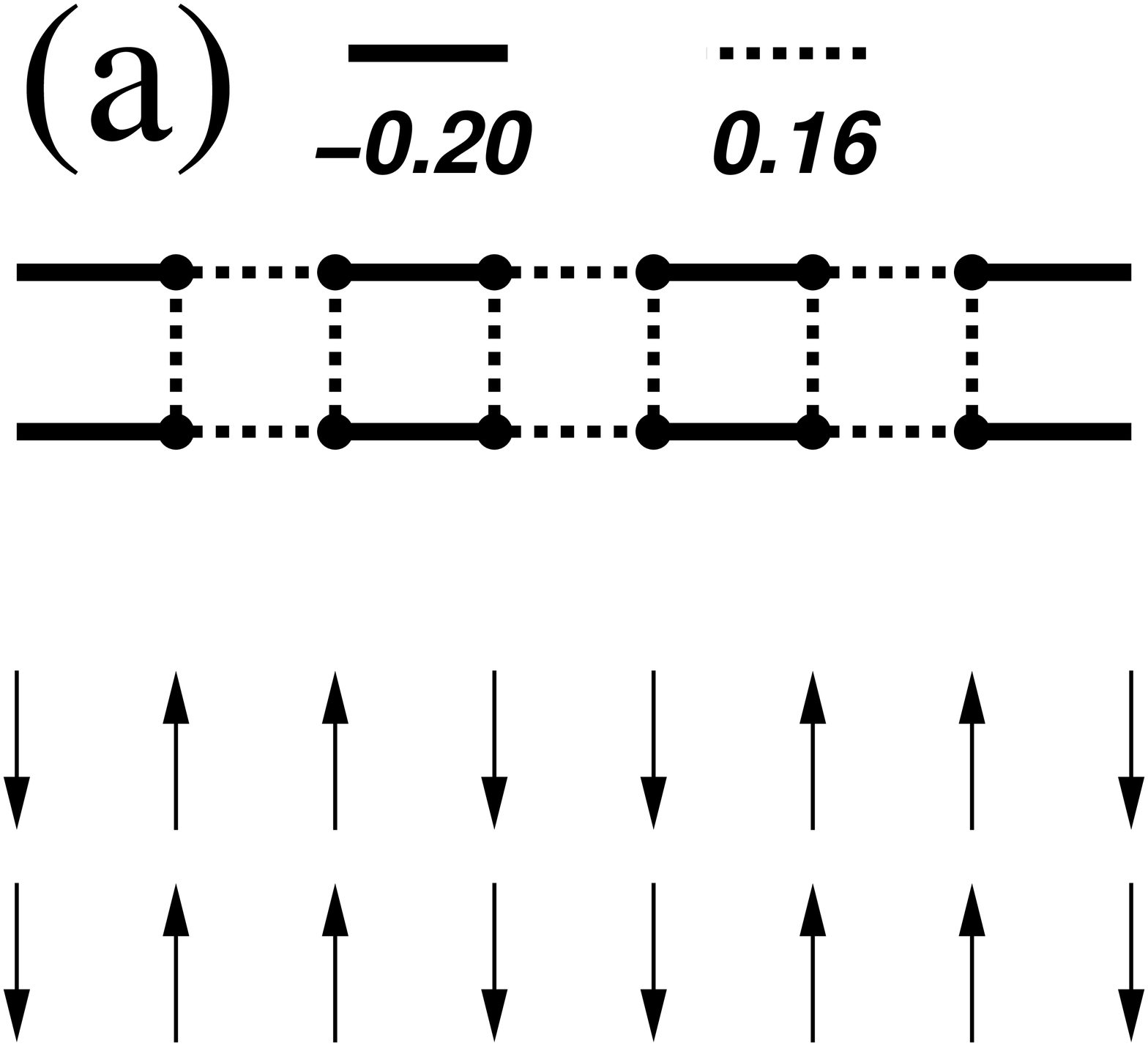} \hspace*{0.5cm} \includegraphics[%
  scale=0.13]{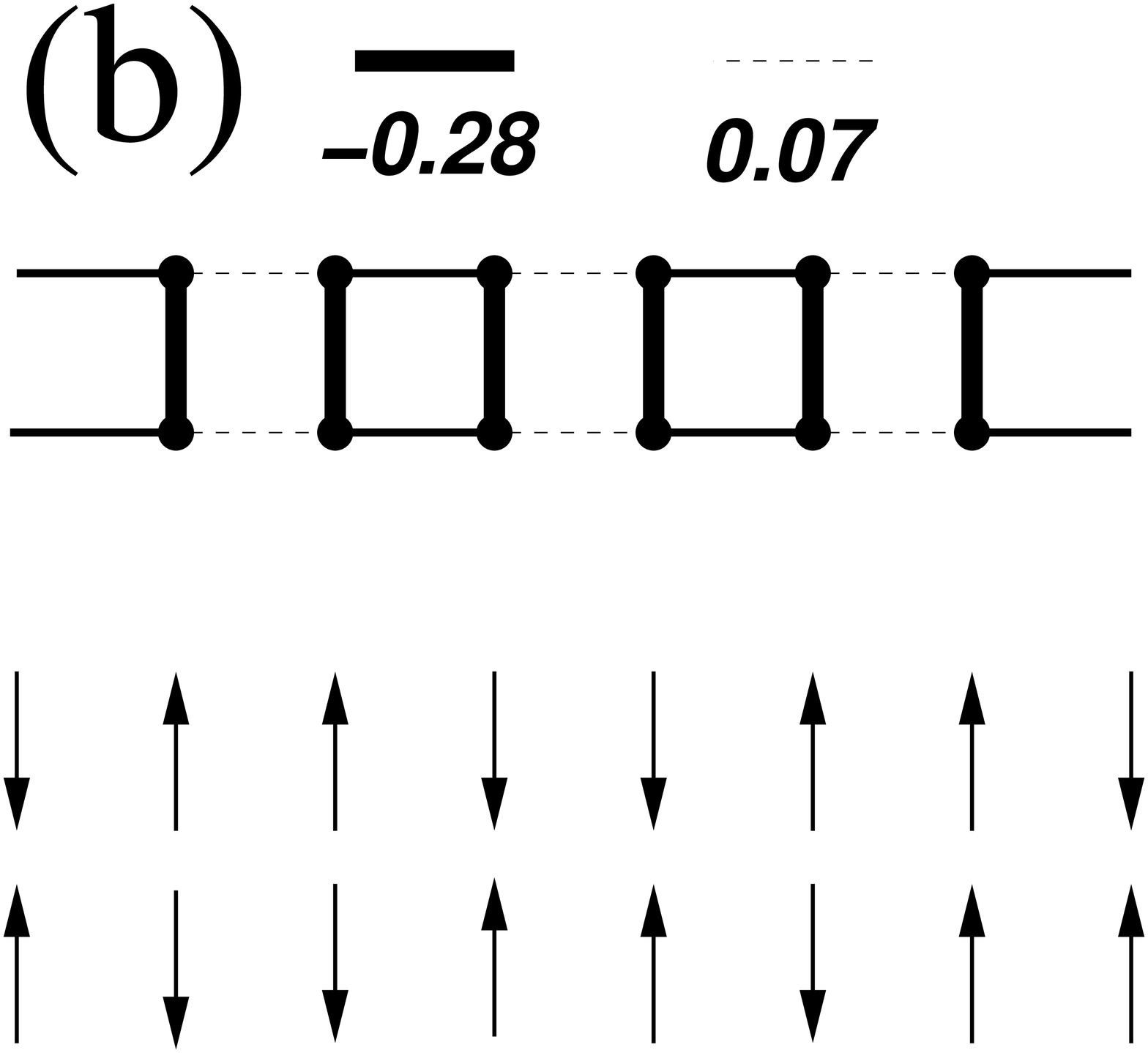}\end{center}

\begin{center}\includegraphics[%
  scale=0.2]{fig3c.eps}\end{center}

\caption{\label{fig3} (a) Nearest-neighbor spin correlations of
the 2-LKL for $L$=$32$, $J$=$0.8$, and $n$=$1/4$. Solid and dashed
lines represent AFM and FM correlations, respectively. The thickness
of the lines is proportional to the magnitude of the correlations.
Only the ladder central portion is presented. Below the correlations,
a classical configuration compatible with them is shown. (b) Same
as (a) but for $n$=$1/2$; (c) Dimer order-parameter $D\left(L/2-1\right)$vs.
$1/L$ for $J$=$0.8$ and $n$=$1/4$.}
\end{figure}

To probe the paramagnetic phase, we calculated the Fourier
transform of the spin-spin correlation function (the spin structure
factor) $S\left(\vec{q}\right)=\frac{1}{2L}\sum_{\vec{r}_{1},\vec{r}_{2}}e^{\vec{q}\cdot\left(\vec{r}_{1}-\vec{r}_{2}\right)}\left\langle \mathbf{S}_{\vec{r}_{1}}^{T}\cdot\mathbf{S}_{\vec{r}_{2}}^{T}\right\rangle $,
where $\mathbf{S}_{\vec{r}_{1}}^{T}$=$\mathbf{S}_{\vec{r}_{1}}+\mathbf{s}_{\vec{r}_{1}}$.
We observed that the maximum of $S\left(\vec{q}\right)$ can appear
in three distinct positions, as indicated in Fig.~\ref{fig1}(a).
For small values of $J$, the maximum of $S\left(\vec{q}\right)$
is located at $\vec{q}$=$(2n,0)\pi$ for $n\lesssim0.5$, while for
$n\gtrsim0.5$ it is at $\vec{q}$=$(n,1)\pi$. 
As an example, in Figs.~\ref{fig2}(a)
and (b) the spin structure factor $S\left(\vec{q}\right)$
is presented for 
the 2$\times$32 cluster with $J$=$0.8$ and densities $n$=$1/4$ 
and $n$=$7/8$,
respectively. These results do not seem to be caused by finite-size effects,
as shown in Fig.~\ref{fig2}(c): the peak becomes more pronounced
as the length $L$ increases. Previous studies of the two-leg Hubbard
model close to half-filling also found that the peak of $S\left(\vec{q}\right)$
appears at $\vec{q}$=$(n,1)\pi$.\cite{noacketal} We have also observed
that, as the density decreases from $n$=$1$, the peak at $\vec{q}$=$(n,1)\pi$
decreases, while another peak at $\vec{q}$=$(2n,0)\pi$ starts to increase,
such that at $n\approx0.5$ they have the same magnitude. In contrast,
for large values of $J$, $S\left(\vec{q}\right)$ is one order of
magnitude smaller and has a small cusp at $\vec{q}$=$(2n,0)\pi$, as
can be seen in Fig.~\ref{fig2}(d). 

Spin dimerization of the localized spins has been detected in the 1D KLM
for both $J$<$0$\cite{garciaetal} and $J$>$0$.\cite{dimer} It would
be very interesting if the dimerization also survives in the 2-LKL,
as this would suggest that it may also be present in the 2D system.
Indeed, we have observed spin correlations in the 2-LKL that resemble
the dimerization of the 1D KLM. In Fig.~\ref{fig3}(a), we show
the spin-spin correlations $\left\langle \mathbf{S}_{\vec{r}_{1}}^{T}\cdot\mathbf{S}_{\vec{r}_{2}}^{T}\right\rangle $
for nearest-neighbor sites of the 2-LKL at density $n$=$1/4$, $J$=$0.8$
and $L$=$32$. The solid (dashed) lines indicate that $D(j)$ is negative
(positive) and the line thickness is proportional to the amplitude
of the correlations. As can be seen, along the legs the dimer order
parameter 
$D(j)$=$\left\langle \mathbf{S}_{(1,j)}^{T}\cdot\mathbf{S}_{(1,j+1)}^{T}\right\rangle $
oscillates with period 2, while the rungs exhibit FM correlations.
We have also found that for $n$=$1/2$, 
$D(j)$ also oscillates with period 2,
as shown in Fig.~\ref{fig3}(b).
However, in this case, the
correlations along the rungs are antiferromagnetic. 
This is not a finite-size effect artifact or caused by open boundaries:
in Fig.~\ref{fig3}(c), the order parameter
at the center of the ladder $D(j=L/2-1)$ vs. $1/L$,
at $J$=$0.8$ $n$=$1/4$, shows a very weak size dependence.
For other densities, more complex spin structures were observed and there
is no analogous simple picture (a similar situation occurs in the 1D KLM away
from quarter-filling\cite{dimer}). For $J$$\gg$$1$ and $0.5$$\lesssim n$$<$$1$,
$\left\langle \mathbf{S}_{\vec{r}_{1}}^{T}\cdot\mathbf{S}_{\vec{r}_{2}}^{T}\right\rangle \sim-10^{-3}$
(for $n$$\lesssim$$0.5$ some small ferromagnetic correlations start to
develop), much less than the values found for small $J$. We also have
verified that, as in the 1D case,\cite{dimer} these spin correlations
can be traced back to the RKKY interaction between localized spins.
This effective, conduction-electron mediated spin-spin interaction
can be obtained from second-order perturbation theory and it is given by
\begin{widetext} 

\begin{eqnarray*}
H_{\rm RKKY} & \sim & J^{2}\sum_{i,j}J_{\rm RKKY}^{1}(|i-j|)(\mathbf{S}_{i}^{1}\cdot\mathbf{S}_{j}^{1}+\mathbf{S}_{i}^{2}\cdot\mathbf{S}_{j}^{2})+2J_{\rm RKKY}^{2}(|i-j|)\mathbf{S}_{i}^{1}\cdot\mathbf{S}_{j}^{2},\end{eqnarray*}

where\[
(J_{\rm RKKY}^{1}(0),J_{\rm RKKY}^{1}(1),J_{\rm RKKY}^{1}(2),...)=\left\{ \begin{array}{ll}
(0,-1,0.66,0.07,-0.41,-0.03,0.28,...) & n=1/4\\
(0,-0.17,1.25,0.73,-1.29,0.10,0.46,...) & n=1/2\end{array}\right.,\]

\[
(J_{\rm RKKY}^{2}(0),J_{\rm RKKY}^{2}(1),J_{\rm RKKY}^{2}(2),...)=\left\{ \begin{array}{ll}
(-1.46,0.23,0.90,0.09,-0.41,-0.03,0.28,...) & n=1/4\\
(4.35,2.30,-2.42,0.07,0.70,0.40,-0.86,...) & n=1/2\end{array}\right.,\]

\end{widetext}and $\mathbf{S}_{j}^{1}$ ($\mathbf{S}_{j}^{2}$) is
the localized spin in the first (second) leg and rung $j$. For 
simplicity, only the first few RKKY couplings were shown.
Let us now focus on density $n$=$1/4$. All $J_{\rm RKKY}^{1}(l)$ have
signs that favor a classical configuration along the legs as $\uparrow\uparrow\downarrow\downarrow\uparrow\uparrow\downarrow\downarrow$.
Note that spin dimerization is expected in a spin chain with first- and
second-neighbor interactions $J_{1}$ and $J_{2}$, if $J_{2}$$>$$0$
and $-4J_{2}$$<$$J_{1}$$<$$0$.\cite{itoiqin} The first two RKKY couplings
$J_{\rm RKKY}^{1}(1)$ and $J_{\rm RKKY}^{1}(2)$ along the legs satisfy this
inequality, and further neighbors couplings favor the classical
configuration. Moreover, the couplings $J_{\rm RKKY}^{2}(l)$ between
legs also favor the predominant ferromagnetic alignment across
the rungs (except for $J_{\rm RKKY}^{2}(1)$=$0.23$, which is nevertheless
small), also in agreement with the classical picture presented in
Fig.~\ref{fig3}(a). A similar analysis also holds for the case $n$=$1/2$.
It is interesting to note that, in this case, the legs are coupled
antiferromagnetically and $J_{\rm RKKY}^{2}(l)$ has larger magnitudes
than at $n$=$1/4$. This fact suggests that at $n$=$1/2$ the legs are
more strongly coupled than at $n$=$1/4$. Indeed, our numerical results
of Figs.~\ref{fig3}(a)-(b) confirm this expectation. Thus, the RKKY
interaction appears to naturally lead to the spin structure shown
in Figs.~\ref{fig3}(a)-(b). It is interesting to note that unusual
ordered spin structures have been observed in some heavy fermion compounds
(see, for example, Ref.~\onlinecite{granadoetal}). Our results suggest
that the effective RKKY interaction may be their origin.

\begin{figure}
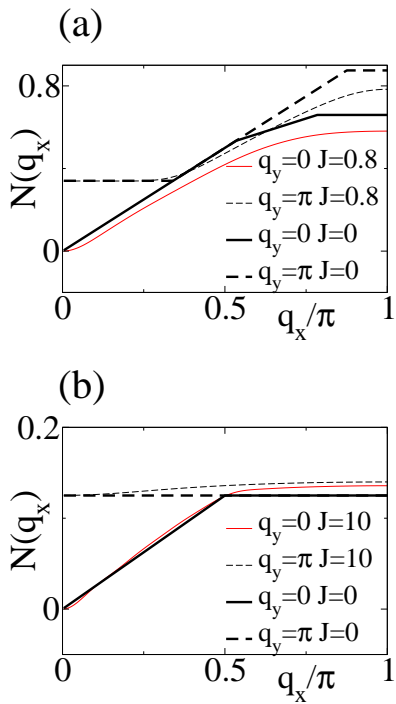

\begin{center}\includegraphics[%
  scale=0.20]{fig4a.eps}\end{center}

\begin{center}\includegraphics[%
  scale=0.20]{fig4b.eps}\end{center}

\caption{\label{fig4} The charge structure factor $N\left(\vec{q}\right)$
vs. $q_{x}$ for the 2$\times$32 cluster and $n$=$7/8$: (a) $J$=$0.8$; 
(b) $J$=$10$.
We also show $N_{0}^{\frac{1}{2}}\left(\vec{q}\right)$ and $N_{0}^{0}\left(\vec{q}\right)$
(see text). }
\end{figure}

We have also calculated the charge structure factor $N\left(\vec{q}\right)$=$\frac{1}{2L}\sum_{\vec{r}_{1},\vec{r}_{2}}e^{\vec{q}\cdot\left(\vec{r}_{1}-\vec{r}_{2}\right)}\left\langle \delta n(\vec{r_{1}})\delta n(\vec{r_{2}})\right\rangle $,
where $\delta n(\vec{r_{1}})$=$n(\vec{r_{1}})-\langle n(\vec{r_{1}}) \rangle$. Previous
work on the 1D KLM has shown that the qualitative 
behavior of $N\left(\vec{q}\right)$
in the extreme weak- and strong-coupling limits could be ascribed to
free spin-$\frac{1}{2}$ and spinless fermions, respectively.\cite{xavier3}
The same analysis clarifies
the behavior of $N\left(\vec{q}\right)$ in the 2-LKL. Let us call
$N_{0}^{S}\left(\vec{q}\right)$ the charge structure factor of free
fermions with spin-$S$ in a two-leg nearest-neighbor tight-binding
ladder.\cite{xavier3} In Fig.~\ref{fig4}(a), $N\left(\vec{q}\right)$ is shown
for the 2-LKL with $L$=$32$, $J$=$0.8$, and $n$=$7/8$, 
as well $N_{0}^{\frac{1}{2}}\left(\vec{q}\right)$.
The behavior of $N\left(\vec{q}\right)$ is fairly similar to free
spin-$\frac{1}{2}$ fermions. On the other hand, 
in the strong coupling limit $N\left(\vec{q}\right)$ approaches the
structure factor $N_{0}^{0}\left(\vec{q}\right)$ of spinless fermions 
(see Fig.~\ref{fig4}(b)).

In conclusion, we have explored the phase diagram of the two-leg Kondo
lattice model away from half-filling. Our results show that a ferromagnetic
phase is present only for small densities, unlike the 1D Kondo chain
but consistent with mean field studies.\cite{lacroixcyrot} We have
found that the charges behave basically as free fermions. On the other
hand, the spins have non-trivial behavior. The
peak of the spin structure factor for small values of $J$ is located
at $\vec{q}$=$(2n,0)\pi$ for $n$$\lesssim$$0.5$, and at $\vec{q}$=$(n,1)\pi$
for $n$$\gtrsim$$0.5$. For large values of $J$ 
and $n$$\gtrsim$$0.4$ $S\left(\vec{q}\right)$
has only a small cusp at $\vec{q}$=$(2n,0)\pi$. We have also shown
that dimerization is present in the 2-LKL at densities $n$=$1/4$ and
$n$=$1/2$, and that the RKKY interaction can tentatively explain this unusual spin
arrangement.

This work was supported by grants from FAPESP (00/02802-7 (JCX) and
01/00719-8 (EM)), CNPq (301222/97-5 (EM)), and DMR (0122523 and 0312333
(ED)).

\end{document}